# The Physics of Interstellar Travel

Coryn Bailer-Jones, Max Planck Institute for Astronomy, Heidelberg
https://bailer-jones.www3.mpia.de/pit.html
November 2026

This is an earlier version of a later-edited article that will appear in *Physik Journal* in December 2026 (https://pro-physik.de/zeitschriften/physik-journal/). It is based on my book *The Physics of Interstellar Travel* (2026), published by Cambridge University Press available open access from https://doi.org/10.1017/9781009700429. No AI tools were used to write this article.

## Introduction

In August 2012, the Voyager 1 spacecraft was 121 au from the Sun and racing away at 17 km/s. It was 35 years since its launch and 32 years since its second and final planetary encounter (with Saturn). As the spacecraft hurtled on through the apparent emptiness of space, its onboard detectors measured a significant decrease in the density of particles from the Sun and an increase in Galactic cosmic rays. Voyager had reached the heliopause, the surface beyond which the interstellar medium (ISM) dominates over the particles from the Sun, considered by some astronomers as the edge of the solar system. By this measure, Voyager 1 had become the first human-made interstellar object.

Over the past 60 years we have sent spacecraft to explore not only every planet in our solar system, but also asteroids, comets, and even distant Kuiper Belt objects. While there is still much to learn about our solar system, it is just one of many. The next logical step is to explore neighbouring star systems.

We already know something about planets around the nearest stars from observations made on Earth and in Earth's orbit. But just as with planets in our solar system, far more could be learned from in-situ observations. The close proximity of a spacecraft will increase the spatial resolution and light gathering power of onboard telescopes by orders of magnitude, allowing us to resolve geological features such as volcanoes and oceans and to look for signs of life such as chemical tracers or vegetation. A spacecraft that goes into orbit could manoeuvre to observe a planet and its moons or rings from different perspectives and illuminations. Other instruments could sample particles - atoms, ions, dust - and measure magnetic and electric fields, things we cannot do remotely at all. And as an interstellar spacecraft will spend most of its time journeying through interstellar medium (ISM), it could study this in situ for the first time.

Yet even our nearest stellar neighbour, Proxima Centauri, is very far away, some 270 000 astronomical units (au), compared to 'just' 30 au to the outermost planet in the solar system, Neptune, at 32 au (1 au = 150 million km is the average Earth-Sun distance). Only a few spacecraft have left the solar system, and the most distant of these, Voyager 1, is 'only' 172 au away (as of November 2026), nearly 50 years after it was launched. It will also cease to operate within a few years from now.

To get to Proxima Centauri in, say, 50 years, a spacecraft will need to move at 25 000 km/s, or 8.5% the speed of light (0.085 c), around 1500 times faster than Voyager 1. This speed cannot be achieved with chemical rockets (see Box 1). Achieving such an enormous speed requires new technologies, such as nuclear fusion rockets or ultralight laser-driven sails. Other challenges await us too: The spacecraft must navigate autonomously to the target, survive high energy impacts from gas and dust in the ISM, and be able to send back data over the enormous distances traversed.

Such an undertaking is still well beyond our capabilities. But the technological developments required to achieve it in, for example, fusion and materials science, will have benefits in other areas of life. The development of an interstellar spacecraft would motivate precursor projects, such as fast missions to explore icy bodies in the outer solar system or the nearby interstellar medium. And while it may take many decades before we can send an uncrewed spacecraft to the nearest stars, the broad aspects of physics required to understand the challenges and identify solutions makes this topic of great pedagogical value (Bailer-Jones 2026).

In this article I look at the key physics behind getting an uncrewed spacecraft to the nearest stars in a human lifetime. The scope is limited to proven, existing physics, so leaves out fun but far-fetched concepts such as warp drives and wormholes.

## Nuclear rockets

The chemical combustion of hydrogen with oxygen releases around 10 MJ of energy per kilogramme of fuel. Nuclear fission and fusion reactions, in contrast, release on the order of $10^8$ MJ/kg. If all of this could be converted into the kinetic energy of a rocket exhaust, it would produce an effective exhaust velocity of about 14 thousand km/s, or 0.05 c ($E/m = v_e^2/2$). The challenge is not so much producing sufficient energy, but rather converting it into a high velocity propellant.

---

## Box 1: The rocket equation

A rocket works through the principle of conservation of momentum: propellant exhausted at a velocity $v_e$ causes the rocket to move in the opposite direction. When the initial mass of the rocket is $M_i$ and the final mass of the rocket is $M_f$ (the mass of the propellant ejected is $M_i - M_f$), the change in velocity of the rocket in deep space (no gravity) is

$$\Delta v = v_e \ln\left(\frac{M_i}{M_f}\right).$$

$v_e$ is measured relative to the rocket and is called the *effective exhaust velocity* because it often consists of particles moving at different speeds. The logarithmic dependence of $\Delta v$ on the mass ratio means the former increases only very slowly. Liquid hydrogen-oxygen propellant provides $v_e \simeq$ 4.5 km/s, about the largest achievable with a chemical propellant. An enormous mass ratio of 100 (most real rockets have less than 20), would deliver only $\Delta v =$ 13.5 km/s. Even when boosted by gravity assists around the planets, as used by Voyager and other missions, our massive chemical rocket would still move at no more than a few tens of km/s relative to the Sun. To get to a high velocity with a rocket, we need to increase $v_e$ by orders of magnitude, something we can do if we replace chemical reactions with nuclear ones.

The rocket equation also explains why braking at the target star to get into orbit is so much more costly than a flyby mission: The rocket would have to provide twice the $\Delta v$, which for a given $v_e$ means we need the *square* of the mass ratio, e.g. 10 000 rather than 100.

---

A straight forward approach is to use a nuclear reactor to heat (but not combust) an inert propellant, such as hydrogen, which then escapes through a converging-diverging nozzle that maximizes the conversion of thermal energy into kinetic energy. Other than the energy source, this is just how a conventional rocket works. Such *nuclear thermal fission rockets* have been tested, for example in the US-American NERVA programme in the 1960s (Robins & Finger 1991). However, as the propellant temperature has to be limited to around 3000 K to avoid melting the fuel and engine, the maximum effective exhaust velocity was no more than about 10 km/s.

A more efficient energy conversion can be attained by using the products of the nuclear reaction directly as the propellant. This is the idea of the *fission fragment rocket*. In contrast to a conventional fission reactor in which the fuel is arranged into rods, the fuel is manufactured as dust-sized grains, thereby allowing the fission products to escape from the unburned fuel. The fission products are too hot to be collimated by a conventional nozzle, but as they are charged, we can instead use a magnetic nozzle. This too converts random thermal motions into an ordered flow, but now with magnetic fields. A specific design by Werke et al. (2012) uses a 1 GW reactor several metres in size with an engine mass of 110 tonnes to produce $v_e =$ 5170 km/s, or 0.017 c. The mass flow rate from such an engine is quite low, however, so even with a large mass ratio it would take a long time to achieve a large $\Delta v$. With a few tens of tonnes of fuel, such a rocket could - in a rather optimistic scenario - get to Proxima Centauri in about a thousand years.

Instead of using a steady reactor, we can extract nuclear energy explosively. One can in principle set off a sequence of small fission bombs below a rocket and, through the use of a thick 'pusher plate' attached to the rocket, absorb the momentum of the explosion fragments to propel the rocket. This idea was actually researched by NASA in the 1960s under the name of Project Orion (Schmidt et al. 2000), and they even performed tests, albeit with conventional explosives. Fortunately, the nuclear test ban treaty led to the project's cancellation.

Nonetheless, the idea of using small nuclear explosions lives on in the idea of the *fusion pulse rocket*. Unlike fission, fusion has no critical mass, so very small explosions can be used. The fusion of deuterium with helium-3

$$\mathrm{D} + {}^{3}\mathrm{He} \;\rightarrow\; {}^{4}\mathrm{He}\ (3.67\,\mathrm{MeV}) + \mathrm{p}\ (14.68\,\mathrm{MeV})$$

is favoured, because both fusion products are charged and so can be coupled to the rocket via a magnetic nozzle. The alternative deuterium plus tritium reaction, currently being explored in the International Thermonuclear Experimental Reactor (ITER) and other projects for terrestrial electricity production, emits a neutron instead of a proton, which cannot be magnetically coupled. The above reaction requires a higher temperature than D+T, however. Another difficulty is that helium-3 is barely present on the Earth, although it could be scraped from the moon's surface, where it is implanted by the solar wind.

A landmark study of a fusion pulse rocket was undertaken by the British Interplanetary Society in the 1970s (Bond & Martin 1978). Starting above the Earth's atmosphere, this involved an enormous two-stage rocket called Daedalus with a total mass of over 50 thousand tonnes (20 times the Saturn V rocket that took the Apollo astronauts to the moon), almost all of which was deuterium/helium-3. This fuel was arranged into small pellets, each 3 gramme in mass (0.3 gramme in the second stage). These were to be compressed and heated with high intensity laser beams to ignite fusion, each pellet releasing 170 GJ of energy that would deliver $v_e =$ 0.033 c. This was to happen 250 per second over four years, giving the rocket a final velocity of 0.123 c. Daedalus was to carry a 450 tonne payload six light years to the second closest star to the Sun, Barnard's Star.

Developing such a rocket requires big leaps in technology. Inertial confinement fusion with lasers has so far yielded four orders of magnitude less energy per pellet than Daedalus planned. Only in 2022 did the National Ignition Facility produce more energy from the fusion (of deuterium and tritium) than was delivered by the laser (although a lot more energy was required to power the laser). It also operated at a rate of just one or two pellets per day.

Fusion energy can be exploited in other ways, such as siphoning off some of the plasma from a reactor to create a continuous thrust, analogous to the fission fragment rocket, or to heat an inert propellant, analogous to the thermal fission rocket. One such concept of the latter is the *direct fusion drive*, a reactor that confines the plasma within a torus using the field-reversed configuration (as opposed to the tokamak in ITER). A study by Cohen et al. (2019) suggests this could achieve $v_e = 0.05$ c using deuterium as the inert propellant. However, the power per unit mass that is being targeted still lies far below that necessary to accelerate a rocket to the high speeds required.

High velocity nuclear rockets still lie far beyond our capabilities. Yet the basic principles are understood, interest in fusion is growing, and many engineering developments are moving in the right direction.

## Laser sails

The big drawback of rockets is that most of their fuel is required to propel the fuel itself. We can overcome this by using an external source of momentum. This is the idea of the solar sail, which uses the pressure of photons from the Sun. The force exerted by perfectly reflected photons is very low, a force per unit power of just $2/c = 6.7$ nN $W^{-1}$, so the pressure from the complete solar spectrum at the distance of the Earth (1361 W $m^{-2}$) is just $9.1\ \mu$Pa. Solar sails must therefore have a very low mass per unit area. If this is as low as 1.5 grammes per square metre - seven times less than plastic food wrap - and the sail is pointed at the Sun, then the outward force from the Sun's radiation equals the inward force from its gravity (at all distances, because both vary with inverse square distance).

Solar sails have been deployed and tested in Earth orbit. These demonstration missions all had relatively high mass per unit area (100-200 times the above-stated values). Future material developments combined with a close approach to the Sun (the 'sundiver') could achieve velocities of tens or even hundreds of km/s. This has numerous applications for solar system exploration, but is not fast enough for interstellar travel.

Replacing the Sun with a high intensity laser achieves much higher accelerations. The relativistic calculations are not complicated, but when the total energy $E_l$ emitted by the laser (and that reaches the sail) is much less than the rest energy ($mc^2$) of the sail, then the final velocity of a perfectly reflecting sail is approximately $2E_l/mc^2$ (e.g. Bailer-Jones 2026, section 11.1.1). Thus to reach 0.085 c, a ten gramme sail would need to receive about 11 GWh of energy from the laser. This might not seem so large until we consider that even the most powerful long-duration lasers in existence have powers of about 1 MW, so would have to operate for nearly 500 days to deliver this energy. With 1 MW, when the sail reaches just 0.01 c it will already have travelled 45 au. But by this distance the laser beam will have spread out significantly due to diffraction: For a 1 micron wavelength laser transmitted by a 10 m diameter telescope, the central part of the diffraction pattern (the Airy disk) of the beam has a diameter of 800 km! This is much larger than our sail, which only has a diameter of a few metres. The vast majority of the laser power will therefore have been lost around the sail edges, so it would not achieve anything close to the required velocity. A larger sail doesn't help here, because it would be more massive and so need even more energy to get to the same speed.

This tells us that, to accelerate even a very low mass sail to 0.085 c, we need a much more powerful laser in order to accelerate it more quickly - on the order of minutes - before most of the power is lost to diffraction. Yet even with a 100 GW laser, we still need a telescope array with an effective aperture on the order of a kilometre if most of the energy is not to be lost. With lower powers, we need an even larger telescope array.

More detailed calculations for laser sails can be found in Parkin (2018), which were made as part of the Breakthrough Starshot Initiative, although ideas of - and publications on - laser sailing go back to soon after the invention of the laser. The required laser power and telescope size are still far beyond what we can currently build today. The sail material must be extremely light, with a highly reflectivity and/or melting temperature. No such material exists, but there have been some positive developments in the laboratory, for example in photonic crystals. Lin et al. (2025) provides a recent review.

The laser sail is the main plausible alternative means of propulsion to the nuclear rocket. A fundamental difference is that laser sails will have to be very light, with a payload mass of perhaps just a few grammes. This severely limits the size and scope of the scientific instruments that can be carried. However, much of the construction cost would go into the huge laser array; the unit cost of each sail will be small in comparison. It therefore makes sense to launch many - perhaps hundreds or thousands - of sails, each of which can carry just one simple instrument and perform just part of the mission. A swarm of multiple spacecraft also offers redundancy against loses.

Unlike a rocket, laser sails have no means of braking at the target, so they would have to perform a flyby, rather than go into orbit. This greatly limits the amount of science that can be done in the few hours that a flyby takes, even with a swarm.

## Surviving the interstellar medium

The interstellar medium is filled with particles. The density of the ISM varies by orders of magnitude across the Galaxy, but near to the Sun it is around one particle per cubic centimetre, predominantly hydrogen atoms. There are also dust particles of a range of sizes, and although they are outnumbered by the gas atoms by a factor of $10^{12}$, their mass per unit volume is only 100 times less.

As our spacecraft moves through the ISM, it will collide with many atoms and dust particles. An atom will be completely ionized by its impact, with the kinetic energy of the resulting ion dissipated in the spacecraft material which can result in permanent damage via melting. The penetration depth depends on several factors and ranges over orders of magnitude, but is of order millimetres. The electrons from the impacting atom produce significant amount of damaging X-rays through the process of bremsstrahlung.

Dust impacts are much rarer, but their larger masses makes them a bigger hazard overall. Experiments at velocities of 0.05-0.1 c have not been done, but simulations by Hoang et al. (2017) indicate that a surface layer of about 1 cm of graphite or quartz shielding should protect the spacecraft from all but the largest dust impacts. Most of the area and much of the depth of this shield would be eroded over the journey. Very large dust grains will be impossible to shield against and could destroy a spacecraft, although such impacts are less likely. A swarm of spacecraft like that mentioned earlier for laser sails is one way to mitigate the mission risk.

Because most impacts arrive in the direction of travel, interstellar spacecraft will be long and thin to lower the probability of collisions.

We might be able to decelerate the spacecraft via a drag on the ISM's charged particles. An electric sail is a large grid of positively charged wires that repel the positively charged ions in the ISM. A magnetic sail achieves something similar by setting up a magnetosphere around the spacecraft using a current run through a superconducting loop. Calculations show that both can achieve a significant deceleration, although they require huge potentials or currents. The structures are tens or hundreds of kilometres in size, and even with not-yet-existent light materials they would have masses of tonnes. So they are only usable with already massive spacecraft such as nuclear rockets, but not with sails.

## Communication

When planetary spacecraft send their data back to the Earth via radio signals, diffraction broadens the transmitted beam significantly, resulting in an extremely weak signal at the Earth. After New Horizons encountered Pluto in 2015 at a distance of 32 au from the Sun, it transmitted its data using a 2.1 m-diameter antenna powered by two 12 W transmitters (Fountain et al. 2008). These signals were received by a 70 m-diameter antenna. This link could achieve a data rate of only 1 to 2 kb/s, and took 15 months to transmit 6 GB of data.

Proxima Centauri is 8000 times further away than Pluto, so to reduce the diffraction loses, communication will probably be done using visible or near-infrared lasers instead of the radio. Optical communication has already been tested in space, for example by the DSOC experiment on NASA's Psyche satellite, which sent signals over 3 au to the Earth (Biswas et al. 2024).

Using the Friis transmission equation (Box 2), we can compute the power received at a telescope on Earth in terms of the size, distance, and wavelength of the transmitter. Table 1 shows that when a 1 m-diameter transmitter with 10 W power at Proxima Centauri sends signals at 1 $\mu$m wavelength, a 10 m receiver at the Earth will receive only two photons per second. As mass - and therefore transmitter and power - on a spacecraft is limited, we would use a much larger array of receivers to permit a larger rate.

---

## Box 2: Friis transmission equation

When a transmitter of area $A_t$ and power $P_t$ transmits a signal at wavelength $\lambda$, a receiver of area $A_r$ a distance $r$ away receives a power of

$$P_r = \frac{P_t A_t A_r}{\lambda^2 r^2}$$

to within a numerical factor. This represents an ideal case and assumes no absorption along the path (e.g. in the Earth's atmosphere) or losses in the optics or electronics. The inverse square dependence on $\lambda$ and $r$ encourages the use of short wavelengths for communication over interstellar distances. In practice this mean visual wavelengths, because the technology does not exist for communication at shorter wavelengths, and X-rays require a lot of energy to generate and are hard to collimate. The rate at which photons are received is given by

$$\dot{n}_s = \frac{P_r \lambda}{hc} .$$

---

Table 1: Values from the Friis transmission equation, with terms defined in Box 2. N is the number of receivers, combined incoherently. The first row corresponds to New Horizons transmitting from Pluto. Adapted from Bailer-Jones (2026), CC-BY-NC 4.0.

| $P_t$ [W] | $d_t$ [m] | $d_r$ [m] | $N$ | $\lambda$ | $r$ | $P_r$ [W] | $\dot{n}_s\,[\mathrm{s}^{-1}]$ |
|---:|---:|---:|---:|---:|---:|---:|---:|
| 24 | 2.1 | 70 | 1 | 3.6 cm | 32 au | $1.1 \times 10^{-17}$ | $1.9 \times 10^{6}$ |
| 10 | 1 | 70 | 1 | 3.6 cm | 4.25 ly | $1.4 \times 10^{-26}$ | $2.6 \times 10^{-3}$ |
| 10 | 1 | 10 | 1 | $1\,\mu\mathrm{m}$ | 4.25 ly | $3.8 \times 10^{-19}$ | $1.9 \times 10^{0}$ |
| 10 | 4 | 10 | 100 | $1\,\mu\mathrm{m}$ | 4.25 ly | $6.1 \times 10^{-16}$ | $3.1 \times 10^{3}$ |

We are actually interested not in photons per second, but in bits per second received. The relation between the two depends on the noise. This has many sources, not least the very bright target star which, as seen from the Earth, has a very small angular separation from the spacecraft's transmitter. This starlight could be suppressed using a coronagraph and a narrow band filter. We also want to maximize the number of bits of information that an individual photon can carry. Perhaps surprisingly, by using appropriate modulation techniques such as pulse position modulation (PPM), a photon can carry more than one bit. Laboratory experiments have already achieved 10 to 20 bits per photon (Dacha et al. 2025).

To be able to recover the intended message in the presence of noise requires an error correcting code. Such codes build redundancy into the signal. This increases the length of the signal and so lowers the data rate. The received data rate therefore depends on the noise level, and the level of error in the reconstructed message one is willing to accept.

Several theoretical scenarios have explored what data rates could be achieved for different configurations. The study of Parkin (2020) found that with a 100 W, 4.1 m-diameter optical transmitter, a 30 m-diameter receiver would achieve a data rate of 266 b/s, albeit with an error rate of 1 in 1000. At the other end of the scale, the study of Lubin (2016) assumed a 20 W, 30 m-diameter transmitter, but with an enormous array of receivers equivalent to a telescope 10 km in diameter. This would deliver 390 Mb/s, enough to stream standard definition video from Proxima Centauri.

One of the big challenges of interstellar communication is the enormous lag times, of several years, which preclude coordination between transmitter and receiver. As data will probably have to be sent over many years, some signal loss is inevitable. Error correcting codes that employ long timescale redundancy will be essential to avoid losing critical data.

## Other challenges

There are other challenges to achieving interstellar travel. One is navigation. In the solar system, spacecraft rely on signals from the Earth to determine their position and velocity in some reference frame. An interstellar spacecraft must navigate autonomously, using only signals from natural sources. There are at least two ways to do this. One is through measuring the angular separations between pairs of nearby stars. As a spacecraft moves, these separations change because of parallax and astronomical aberration. If we know the positions of the stars in the reference frame - and we do thanks to ESA's Gaia survey satellite - then changes in these angular separations can be used to locate the spacecraft and determine its velocity (Bailer-Jones 2021).

Another way of navigating is to use the regular X-ray pulses received by rapidly spinning neutron stars (pulsars). The difference in the arrival times of pulses compared to their expected arrival times at the Sun allows the spacecraft to determine its position in one dimension; three pulsars give the three-dimensional position. While this requires a sufficiently accurate initial position of the spacecraft, experiments done by the Chinese space agency in Earth's orbit have demonstrated a positional precision of a few tens of kilometres (Huang et al. 2019).

A particular challenge for gramme-scale sailcraft will be the onboard power supply. Stars are too faint for photovoltaics, batteries have far too low energy densities, and conventional radioisotope thermal electric generators, such as those used on Voyager and New Horizons, are too massive. The energy of radioactive decay can be converted into electricity in semiconductors in betavoltaic devices, but so far those with masses of many grammes produce less than a mW of power.

## The way forward

The most promising propulsion methods for interstellar travel are fusion rockets and laser sails. Although physically possible, both of these propulsion methods require significant technological developments before they can be used. Laser sails would be individually very light and very limited in their capabilities, but as their unit cost is relatively low, we would send a swarm of them to achieve the scientific objectives collectively. Fusion rockets are much more massive, allowing them to carry more substantial payloads, and providing the potential to decelerate, albeit at the cost of having to carry a lot more fuel. Decelerating to go into orbit is nonetheless far more attractive than a flyby mission, because it extends the observing time by orders of magnitude.

With so many unknowns, we are not going to achieve interstellar flight in one leap. There are ideas for precursor missions to the outer solar system to explore the Kuiper Belt and nearby ISM. These could test propulsion, communication, and navigation technologies, as well as characterize the particles in the ISM that a future spacecraft will encounter.

Sending an uncrewed spacecraft to the nearest stars will be an enormous undertaking, costing a lot of time, effort, and money. The scientific benefits would be significant, and there are other benefits in terms of technological development, education, international cooperation, and cultural inspiration. But there are also ethical issues with the venture, not least if we find out that the target system is inhabited. There may also be dangers. Some of the technologies we need are dual use, most obviously a hugely powerful laser, which could be used by a malevolent actor as a weapon. Furthermore, anything moving at 10% the speed of light can be interpreted as a weapon. What might aliens make of an entire swarm headed at them?

Finally, much of the challenge of interstellar travel is predicated on the desire to go fast, to get there within a human lifetime. Is this really necessary? Eventually our ancestors could reap the benefits of what we do now, just as we reap the benefits of our ancestors' activities. Longer term, multi-generational thinking might offer an easier solution.